\documentclass[usenatbib]{mn2e}
\usepackage{graphicx,natbib,times,amssymb}
\def\dnu{$\nu_{max}$}
\def\ddel{$\Delta\nu$}
\def\dhz{$\mu$Hz}
\def\dsunm{$\mathrm{M}_{\odot}$}

\title[characteristics of solar-like oscillations of clusters ]
{Characteristics of solar-like oscillations of clusters simulated by
stellar population synthesis}
\author[Yang et al.]{Wuming Yang$^{1,2}$\thanks{E-mail:
yangwuming@ynao.ac.cn; woomyang@gmail.com},
Zhongmu Li$^{3,4}$, Xiangcun Meng$^{2}$ and Shaolan Bi$^{1}$ \\
$^{1}$Department of Astronomy, Beijing Normal University, Beijing 100875, China.\\
$^{2}$School of Physics and Chemistry, Henan Polytechnic University,
Jiaozuo 454000, Henan, China. \\
$^{3}$Institute for Astronomy and History of Science and Technology,
Dali University, Dali 671003, China.\\
$^{4}$National Astronomical Observatories, Chinese Academy of
Sciences, Beijing 100012, China.}

\begin{document}

\date{ }

\pagerange{\pageref{firstpage}--\pageref{lastpage}} \pubyear{2010}

\maketitle

\label{firstpage}

\begin{abstract}
Using a stellar population-synthesis method, we studied the
distributions of \dnu{} and \ddel{} of simulated clusters with
various ages and metallicities. Except for the confirmed peak (RC
peak) of \ddel{} of red-clump (RC) stars, i.e. core-helium burning
stars, there are a gap and a main sequence (MS) peak in the
distributions of \dnu{} and \ddel{} of young clusters. The gap
\textbf{corresponds mainly} to the Hertzsprung gap phase of
evolution. The RC peak is caused by \textbf{the fact} that the
radius of many RC stars near the zero-age horizontal branch
concentrates in a certain range. The MS peak \textbf{also results
from the fact} that many MS stars which are located in a certain
mass range have an approximate radius in the early phase of MS. The
MS peak barely exists in the simulated clusters with age $\lesssim$
1.0 Gyr. The location of \textbf{the} MS peak moves to a lower
frequency with increasing age or metallicity, which may be applied
to constrain the age and metallicity of young clusters. For the
simulated clusters with Z = 0.02, the frequency of \textbf{the}
location of \textbf{the} dominant RC peak increases with age when
age $<$ 1.2 Gyr, \textbf{and} then decreases with age when age $>$
1.2 Gyr; but \textbf{it} scarcely varies when age $>$ 2.4 Gyr. This
is relative to the degeneracy of \textbf{the} hydrogen-exhausted
core at the time of helium ignition. \textbf{In addition,} the RC
peak is not sensitive to the metallicity, especially for the
clusters with age $>$ 2.4 Gyr. \textbf{Asteroseismical} observation
for clusters with age $\lesssim$ 2.4 Gyr may aid in testing the
theory of the degeneracy of \textbf{the} hydrogen-exhausted core.
Moreover, for the clusters with 1.1 \dsunm{} $< M_{hook} <$ 1.3
\dsunm{}, there are a MS gap and a peak on the left of the MS gap in
the distributions of \dnu{} and \ddel{}, which may be applied to
constrain the central hydrogen abundance of stars in the MS gap and
the peak.

\end{abstract}


\section{Introduction}
Each low-degree p-mode of solar-like oscillations carries unique
information about \textbf{the} stellar interior. Thus
asteroseismology has the capability to probe the interior of stars
and to determine the fundamental parameters of individual stars
\citep{ulri86, goug87, chri02, egge06, yang10a}. However, it is more
difficult to extract \textbf{solid} estimates of individual
frequencies than to extract the mean large frequency separation
(\ddel{}) and the frequency of maximum seismic amplitude (\dnu {}).
Fortunately, the \ddel{} and \dnu{} also allow us to determine the
fundamental parameters of stars (mass and radius) to within a few
percent \citep{kjel95, stel09a, stel09b}. Moreover, the property of
frequency separations has been studied by many investigators
\citep{ulri86, goug87, gabr89, roxb00, yang07, yang09}.
Asteroseismic parameters \ddel{} and \dnu{} will be the primary
parameters for asteroseismology. Using the data observed by the
\emph{COnvection ROtation and planetary Transits (CoRoT)}
\citep{bagl06}, \citet{hekk09} and \citet{moss10} extracted the
\dnu{} and \ddel{} of more than 900 red giant stars. By making use
of the \dnu{} and \ddel{}, \citet{kall10} and \citet{moss10} derived
information on the mass and radius of the oscillating stars.
\textbf{In addition}, the distributions of the \dnu{} and \ddel{}
\textbf{have been} applied to test stellar population synthesis
models \citep{migl09, yang10b} and physical processes such as mass
loss and binary interactions \citep{yang10b}. Asteroseismology has
significantly advanced the theory of stellar structure and
evolution.

Since stars in a cluster are believed to have uniform age and
chemical compositions, research on solar-like oscillations in
cluster stars as a uniform ensemble will strengthen our ability to
\textbf{constrain} stellar models and improve our understanding of
stellar evolution and interior physical processes \citep{stel10b,
gill10}. This potential has led to many attempts to detect
solar-like oscillations in clusters \citep{gill93, gill08, edmo96,
stel07, fran07, stel09}. \textbf{Strong} evidence for \textbf{the}
excess power of oscillations \textbf{has been} found in \textbf{the}
red giants of a few clusters, such as open cluster M67
\citep{stel07}, globular cluster 47 Tucanae \citep{edmo96} and NGC
6397 \citep{stel09}, \textbf{but} oscillation frequencies were not
claimed by \textbf{these} authors. Thanks to the \emph{Kepler}
mission \citep{koch10}, there are four open clusters in the
\emph{Kepler} field of view, NGC 6791, NGC 6811, NGC 6866 and NGC
6819 \citep{gill10}, which span a range in metallicity and age
enclosing solar values. Solar-like oscillations have been clearly
detected \textbf{by \emph{Kepler} Mission} in a large sample of red
giants in the cluster NGC 6819 \citep{stel10a}. The values of \dnu{}
and \ddel{} for the cluster stars were extracted firstly by
\citet{stel10a}. In future observations, oscillations in subgiants,
i.e. Hertzsprung gap stars, and main sequence (MS) stars in these
clusters \textbf{could} be measured by \textbf{the} \emph{Kepler}
Mission \citep{stel10a, gill10, stel10b}. Moreover, the
\emph{PLAnetary Transits and Oscillations of stars mission}
(\emph{PLATO}) \citep{roxb07, cata09} will monitor more than 100,000
bright ($\mathrm{m_{V}}\leq$ 11) stars of all spectral types, which
\textbf{can provide} a large sample of stellar oscillations. Having
seismic parameters among a large sample of cluster stars at various
stages of evolution will allow us to have independent constraints on
the cluster age\textbf{ and test} the various physical processes
that govern stellar evolution \citep{gill10}. \textbf{Because} the
observational accuracy for \textbf{\dnu{} and \ddel{} is very high}
(for example, the uncertainty of \emph{CoRoT} is about 0.75 \dhz{}
for \ddel{} of giant stars \citep{moss10}) and seismic data do not
suffer from uncertainties in distance or extinction and reddening,
asteroseismology could \textbf{thus} provide more stringent
constraints on the cluster age than classical approaches.

In this paper, by making use of a stellar population-synthesis
method, we studied the characteristics of distributions of
\textbf{the} \dnu{} and \ddel{} of clusters with various
metallicities and different ages. We used the Hurley rapid evolution
codes \citep{hurl00, hurl02} to construct stellar models for stellar
population synthesis, \textbf{and} we used Eggleton's stellar
evolution code \citep{egg71, egg72, egg73}, \textbf{as} updated by
\cite{han94} and \cite{pols95, pols98}, to obtain the details of a
model, such as the central hydrogen abundance. The Hurley codes can
rapidly reproduce stellar models computed \textbf{by} using the
Eggleton's stellar evolution code. The equation of state of
\cite{eggl73} \textbf{as} modified by \cite{pols95}, OPAL
\citep{roge92} and \cite{alex94} opacity tables, and standard
mixing-length theory are used in the Eggleton's code. The value of
\textbf{the} mixing-length parameter $\alpha$ and convective
overshooting parameter $\delta_{ov}$ \citep{pols98} is set as 2.0
and 0.12, respectively. \textbf{However, element} diffusion for
helium and metals is not taken into account. \textbf{In addition},
the theoretical \dnu{} and \ddel{} were calculated \textbf{by} using
scaling equations \citep{brow91, kjel95}
\begin{equation}
    \nu_{max}=3050 \frac{M/M_{\odot}}{(R/R_{\odot})^{2}
    \sqrt{T_{eff}/5777 K}}\mathrm{\mu Hz}\,,
    \label{eq1}
\end{equation}
   and
   \begin{equation}
    \Delta \nu=134.9 \frac{(M/M_{\odot})^{1/2}}{(R/R_{\odot})^{3/2}} \mathrm{\mu
    Hz}\,. \label{eq2}
   \end{equation}
The accuracy of these estimates is good to within 5\% for a star
given mass, radius and effective temperature \citep{stel09a}.

The paper is organized as follows. We show our stellar
population-synthesis method in section 2. \textbf{We present the}
results in section 3 \textbf{and discuss and summarize them} in
section 4.

\section{Stellar population synthesis}

To simulate the distributions of \dnu{ } and \ddel{ } of clusters
with different metallicities and various ages, we have performed
detailed Monte Carlo simulations for \textbf{a} binary-star stellar
population (BSP). The basic assumptions for the simulations are as
follows. The lognormal initial mass function (IMF) of \cite{chab01}
is adopted. Firstly, we generate the mass of the primary, $M_{1}$,
according to the IMF. \textbf{For simplicity,} the ratio ($q$) of
the mass of the secondary to that of the primary and eccentricity
(e) are assumed to be a uniform distribution within 0-1.
\textbf{The} mass of the secondary star is \textbf{then determined}
by $qM_{1}$. We assume that all stars are members of binary systems
and that the distribution of separations is constant in $\log a$ for
wide binaries and falls off smoothly at close separation:
\begin{equation}
an(a)=\left\{
 \begin{array}{lc}
 \alpha_{\rm sep}(a/a_{\rm 0})^{\rm m} & a\leq a_{\rm 0},\\
\alpha_{\rm sep}, & a_{\rm 0}<a<a_{\rm 1},\\
\end{array}\right.
\end{equation}
where $\alpha_{\rm sep}\approx0.070$, $a_{\rm 0}=10R_{\odot}$,
$a_{\rm 1}=5.75\times 10^{\rm 6}R_{\odot}=0.13{\rm pc}$ and
$m\approx1.2$. This distribution implies that the number of wide
binary systems per logarithmic interval is equal, and that
approximately 50\% of the stellar systems are binary systems with
orbital periods less than 100 yr \citep{han95}. With these
assumptions, we calculated the evolutions of 5$\times 10^{4}$
binaries with $M_{1}$ between 0.8 and 5.0 \dsunm{} to obtain their
mass, radius and effective temperature. The \dnu{} and \ddel{} of
these stars were calculated using equations (\ref{eq1}) and
(\ref{eq2}).

\section{Calculation results}
Figure \ref{fig1} shows the histograms of \dnu{} and \ddel{} of
simulated clusters with the same metallicity (0.02) but different
ages. Stars hotter than the red edge of the instability strip
\citep{cunh02} are generally not expected to exhibit solar-like
oscillations. The red edge locates roughly between 6500K and 7000K
in our models. The characteristics of theoretical acoustic modes of
these stars were studied by \cite{auda94} and \cite{stel09a}. They
showed that the theoretical large separation $\Delta\nu$ of these
stars could be obtained from their mean density with an uncertainty
of about 10 per cent. Some stars on the left of the red edge may
oscillate as rapidly oscillating Ap (roAp) stars \citep{cunh02,
balo10}. \textbf{A} pattern of nearly equally spaced frequencies and
low frequencies may exist in these stars \citep{balo10}. In our
sample, these stars occur mainly in clusters with age $\lesssim$ 2.0
Gyr \textbf{and} are not discarded. \textbf{However,} in order to
distinguish them from stars expected to exhibit solar-like
oscillations, they are shown in black in Figs. \ref{fig1},
\ref{fig2}, \ref{pz3}, \ref{pz08} and \ref{fmsp}. Moreover, the
histograms of \ddel{} of red-clump (RC) stars, i.e.
core-helium-burning stars, first giant branch (FGB) stars, subgiants
and MS stars of the clusters with age = 1.0 and 5.0 Gyr are plotted
in Figs. \ref{fig2} and \ref{fig3}, respectively.

\subsection{Gaps}
\begin{figure}
     \includegraphics[angle=-90, width=7.5cm]{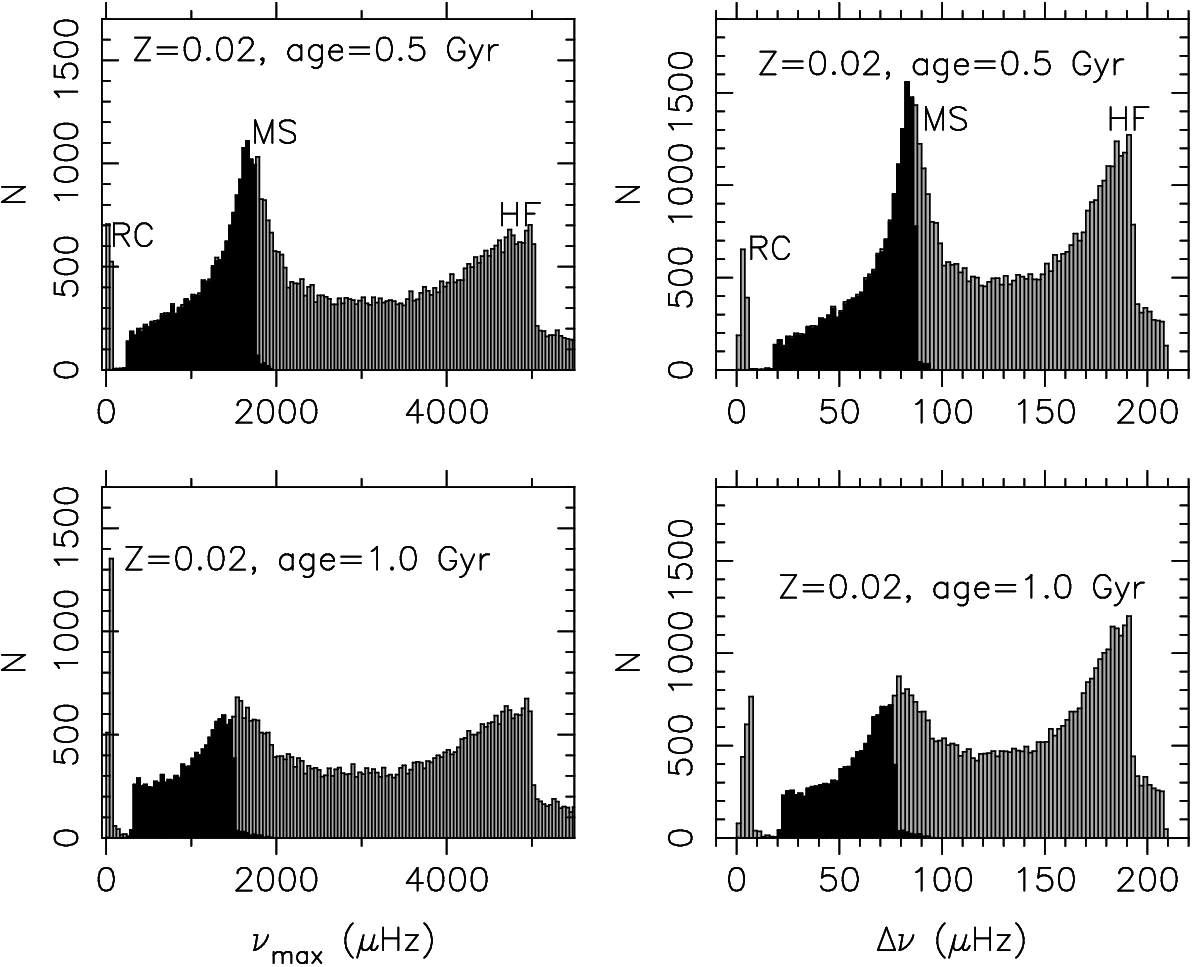}
     \includegraphics[angle=-90, width=7.5cm]{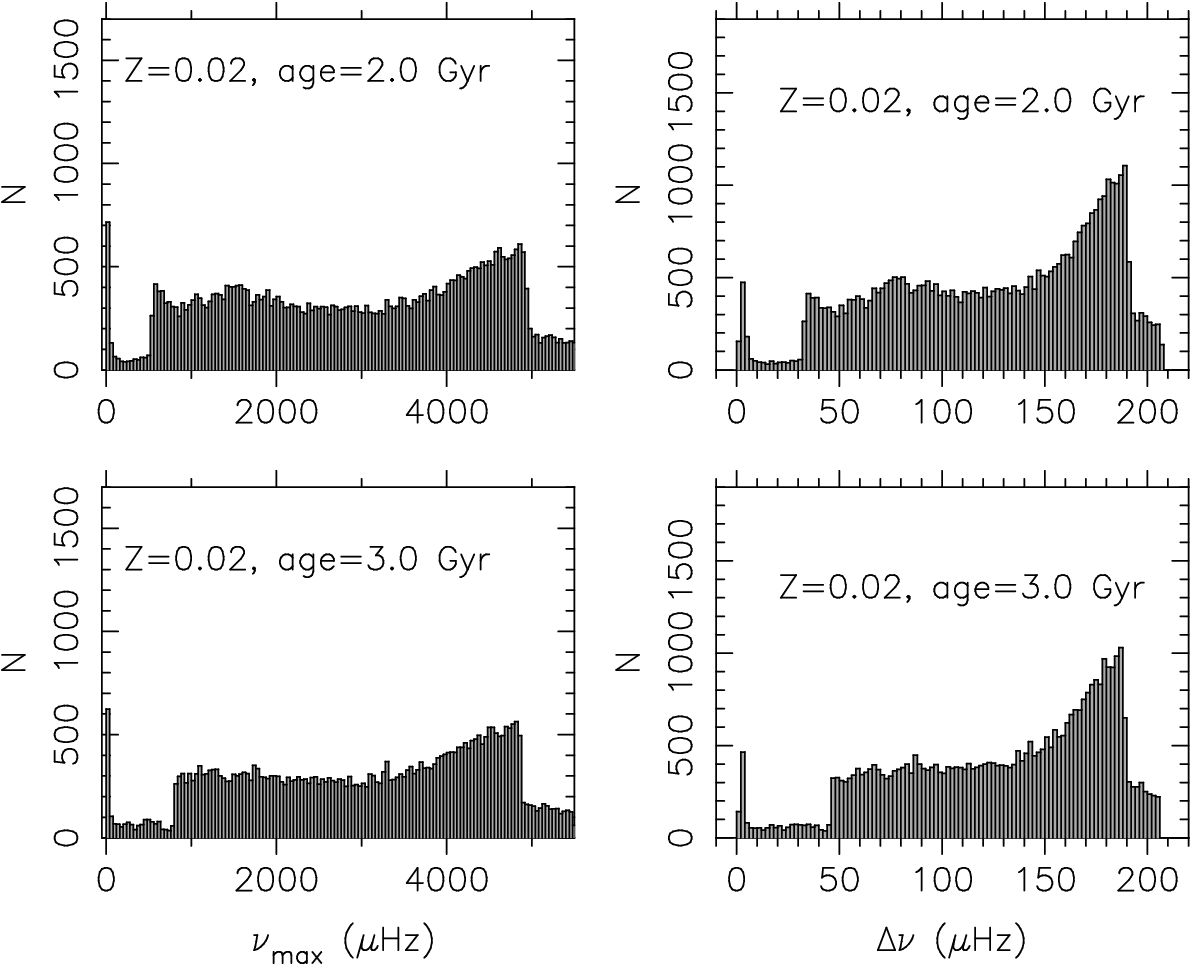}
     \includegraphics[angle=-90, width=7.5cm]{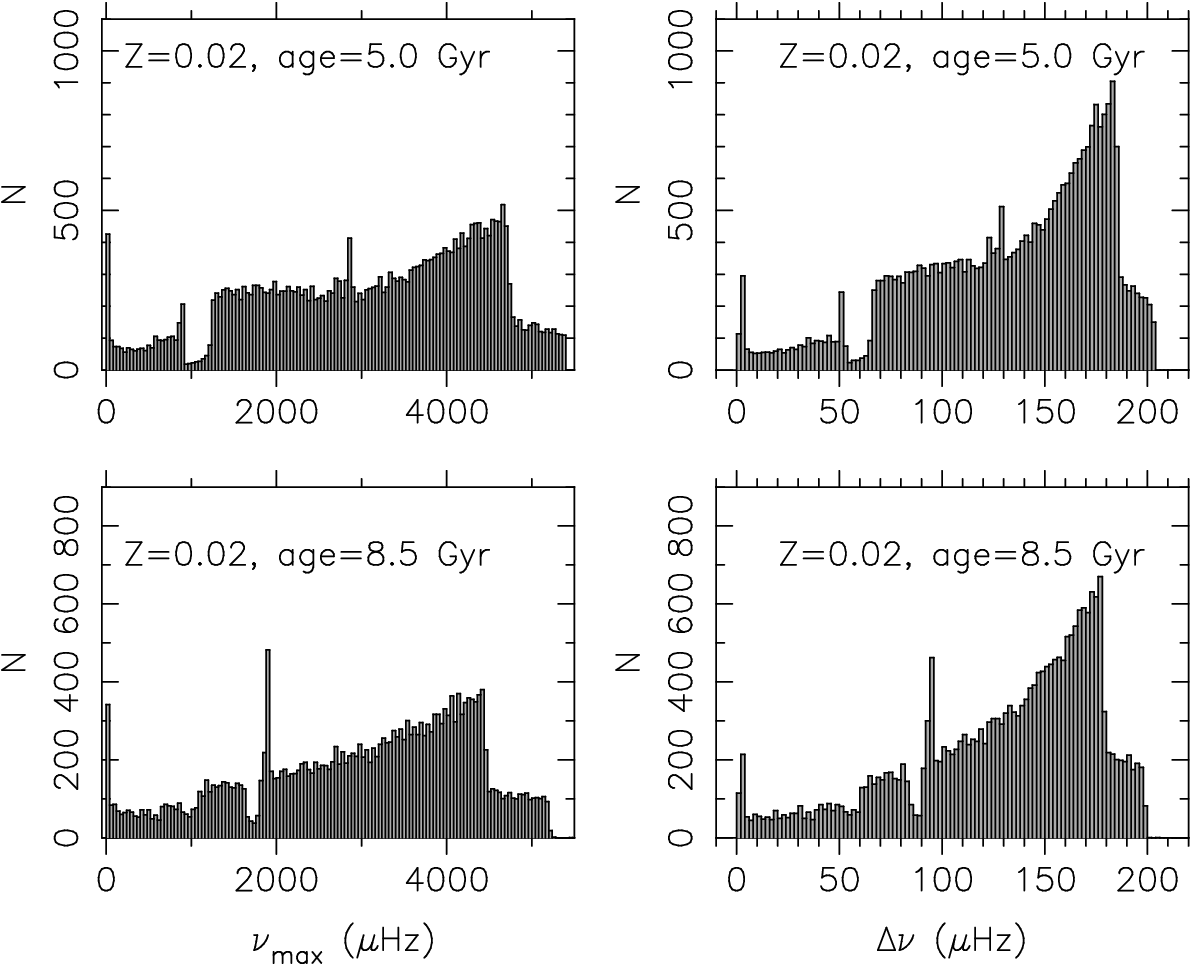}
     \centering
     \caption{The histograms of \dnu{} and \ddel{} of simulated
      clusters with the same metallicity but different ages. Models
      hotter than the red edge of instability strip are shown in black,
      while cooler models are shown in light gray.}
       \label{fig1}
   \end{figure}

Figure \ref{fig1} shows that the distributions of \dnu{} and \ddel{}
of young clusters are obviously different from those of old
clusters. A notable characteristic of the distributions is that
there is a gap (RC-MS gap) between the histogram of RC stars and
that of MS stars. This gap becomes wider with increasing age. It
nearly disappears when age $>$ 5 Gyr.

\begin{figure}
     \includegraphics[angle=-90, width=7.5cm]{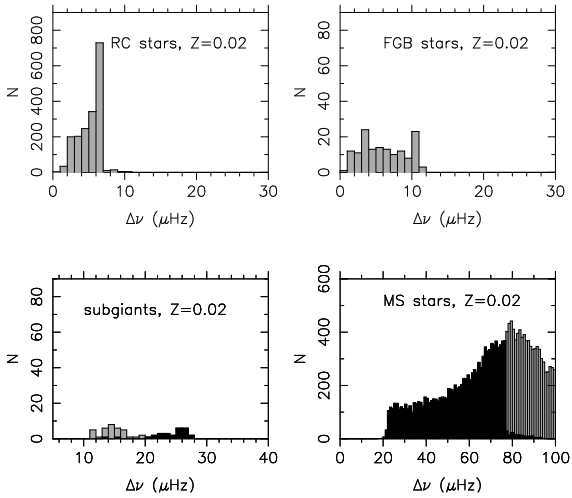}
     \centering
     \caption{The histograms of \ddel{} of RC stars, FGB stars,
      subgiants and MS stars of the simulated cluster with
      age = 1.0 Gyr. Black represents models hotter than
      the red edge of instability strip, while light gray shows cooler
      models.}
     \label{fig2}
   \end{figure}

\begin{figure}
     \includegraphics[angle=-90, width=7.5cm]{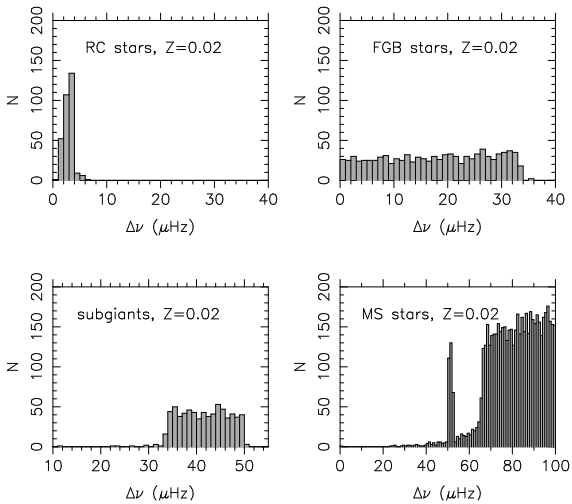}
     \centering
     \caption{Same as Fig. \ref{fig2} but for the simulated cluster
     with age = 5.0 Gyr.}
     \label{fig3}
   \end{figure}

\begin{figure}
     \includegraphics[angle=-90, width=7.5cm]{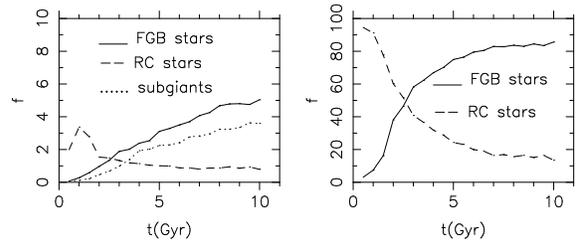}
     \centering
     \caption{The evolution of the fraction f (in percentage)
      of different \textbf{types of} stars of \textbf{the} simulated
      cluster with Z = 0.02.
      The fraction in \textbf{the} left panel \textbf{stands for}
      `FGB over the total number of stars'. \textbf{That} in \textbf{the} right
      panel \textbf{indicates} `FGB/(FGB+RC)'.}
     \label{fig4}
   \end{figure}

In spite of \textbf{this fact,} there are many stars hotter than the
red edge of the instability strip. \textbf{For} the cluster with age
= 0.5 Gyr, the RC-MS gap is located in the range about 6-18 \dhz{}
for \ddel{} and about 80-240 \dhz{} for \dnu{}. For the cluster with
age = 1.0 Gyr, Fig. \ref{fig2} shows that the values of \ddel{} are
basically located in the range about 1-10 \dhz{} for RC stars and
FGB stars, and in the range about 10-30 \dhz{} for subgiants, while
the values are mostly larger than 20 \dhz{} for MS stars. \textbf{In
addition}, the number of subgiants is very small in this cluster,
which leads to the appearance of the gap between about 10 and 20
\dhz{}. For the cluster with age = 5.0 Gyr, Fig. \ref{fig3} shows
that the values of \ddel{} are almost uniformly distributed in the
range \textbf{of} about 1-34 \dhz{} for FGB stars and \textbf{that
of} about 34-50 \dhz{} for subgiants, and that the values are mostly
larger than 50 \dhz{} for MS stars. However, the \ddel{} of RC stars
is almost concentrated in a narrow range of 1-6 \dhz{}. Thus,
although the number of RC stars is less than that of FGB stars and
that of subgiants (see Fig. \ref{fig4}), the RC-MS gap still exists
in the distribution of \ddel{}. \textbf{It} becomes wider \textbf{in
comparison} with that in the cluster with age = 1.0 Gyr. However,
Fig. \ref{fig1} shows that the RC-MS gap of clusters with age $<$
2.0 Gyr can be affected by the stars hotter than the red edge of the
instability strip. If these stars are discarded, the RC-MS gap
should be \textbf{significantly enlarged}. But for the clusters with
age $\gtrsim$ 2.0 Gyr, the distributions of \dnu{} and \ddel{} are
not affected by these stars.

The stars \textbf{located in the} RC-MS gap are primarily subgiants
for young clusters, but the stars \textbf{located in the gap} are
mainly subgiants and FGB stars, \textbf{with a small number} of RC
stars and of MS stars for `middle-age' clusters. The number of
subgiants and FGB stars increases with age, which results in
\textbf{the gradual disappearance of this gap} with increasing age.
The mass of stars in \textbf{an} old cluster is less than that of
the stars in the same evolutionary stage of \textbf{a} younger
cluster, which is partly \textbf{due to a} more significant
mass-loss for old stars, especially for the RC stars. In addition,
\textbf{except the RC stars,} for stars in the same evolutionary
phase \textbf{generally} the lower the mass, the higher the mean
density, i.e. the larger the \ddel{}. Thus \textbf{with regard to}
the values of \ddel{} of subgiants, FGB stars on the bottom of
\textbf{a} FGB and the MS stars at the end of \textbf{a} MS increase
with increasing age. But the values of \ddel{} of RC stars remain in
the range \textbf{of} about 1-10 \dhz{}. Consequently, the RC-MS gap
becomes wider with increasing age.


Moreover, Fig. \ref{fig3} reveals clearly that another gap (MS gap)
exists in the histogram of MS stars. For the cluster with age = 5
Gyr, this gap is located in the range $\sim$ 53-65 \dhz{} for
\ddel{} and $\sim$ 920-1200 \dhz{} for \dnu{}. Our calculations show
that the location of this MS gap changes with age and that it mainly
appears in clusters with age $\gtrsim$ 4.5 Gyr. This MS gap
\textbf{corresponds} to the hydrogen-exhausted phase gap, or MS hook
\citep{hurl00}. For stars with $M >$ 1.1 \dsunm{}, their core is
convective. Owing to mixing in the core there is a sudden depletion
of fuel over a large region, which leads to a rapid contraction of
the core and expansion of \textbf{the} radius on a thermal
time-scale \citep{hurl00}. This causes the appearance of
\textbf{the} MS hook in the Hertzsprung-Russell diagram and the MS
gap in the distributions of \dnu{} and \ddel{}.

There are some sharp edges in the distributions of \dnu{} and
\ddel{} in Figs. \ref{fig1} and \ref{fig2}. For example, the edge
between the histogram of \ddel{} of MS stars and that of subgiants
in the cluster with age = 3.0 Gyr in Fig. \ref{fig1} and the edge of
distribution of RC stars in Fig. \ref{fig2}. For the cluster with Z
= 0.02 and age = 3.0 Gyr, stars with $M <$ 1.45 \dsunm{} are MS
stars, while most stars with $M >$ 1.45 \dsunm{} have evolved into
\textbf{the} MS hook or a later stage. For stars with $M>$ 1.45
\dsunm{}, the time spent in \textbf{the} MS hook is a thermal
time-scale, which is far less than the time of hydrogen burning.
Thus the number of stars in \textbf{the} MS hook \textbf{are few} in
our sample. In addition, the radius of these stars expands very
rapidly in \textbf{the} MS-hook stage, i.e. their mean density
decreases very fast, which leads to a dispersion of their \ddel{} in
a wide range of $\sim$ 20-46 \dhz{}. Therefore, there is a sharp
edge between the histogram of \ddel{} of MS stars and that of
subgiants. For a RC star, its mean density is highest when it is on
zero-age horizontal branch (ZAHB). The evolution of a star from the
top of FGB to ZAHB is very rapid. For the cluster with Z = 0.02 and
age = 1.0 Gyr, the mass of RC stars is \textbf{located almost
solely} in the range $\sim$ 2.1-2.3 \dsunm{}, while stars with lower
mass are mainly in \textbf{the} FGB or an earlier stage. The stars
with mass between $\sim$ 2.1 and 2.2 \dsunm{} are very close to
ZAHB. The values of \ddel{} of these stars are concentrated in the
range $\sim$ 6.0-7.0 \dhz{}. Hence, the distribution of \ddel{} of
RC stars has a sharp edge.

Fig. \ref{fig4} represents the evolution of fractions of subgiants,
FGB and RC stars in a cluster. Our calculations show that the
fractions of FGB and RC stars are scarcely affected by metallicity.
In clusters with age $<$ 1 Gyr, more than 90 per cent of red giants
(including \textbf{only} FGB and RC stars) are RC stars. But over 80
per cent of red giants are FGB stars in clusters with age $>$ 8.0
Gyr. However, the fraction of RC stars is approximately equal to
that of FGB stars in clusters with age $\simeq$ 2.5 Gyr, which is
caused by \textbf{the fact} that for low mass stars the time spent
ascending the FGB is basically comparable with that of \textbf{the}
core helium-burning of \textbf{the} more massive stars in these
clusters.

\subsection{Peaks}
Except for the gaps, Fig. \ref{fig1} shows that there are three
peaks in the distributions of \dnu{} and \ddel{} of young clusters.
One is at `low' frequencies (labeled RC peak), one is at `middle'
frequencies (labeled MS peak), and another is at `high' frequencies
(labeled HF peak). Figs. \ref{fig2} and \ref{fig3} display clearly
that the RC peak \textbf{derives} mainly from RC stars, and that the
MS peak is from MS stars. The HF peak \textbf{derives also} from MS
stars and is caused by IMF. For MS stars with the same age and
metallicity, the lower the mass, the higher the mean density, i.e.
the larger the \dnu{} and \ddel{}. \textbf{In addition}, according
to IMF, the lower the stellar mass, the more the number of stars.
Thus there is a peak at `high' frequencies. In our models, the HF
peak is caused by the stars with $M \sim$ 0.8 \dsunm{}, which is the
lower limit of \textbf{the} initial mass of the primary
\textbf{star} in a primordial binary system. Because the mean
density, i.e. the \ddel{}, of a MS star decreases with age, the
location of \textbf{the} HF peak moves to a lower frequency with
increasing age. The HF and RC peaks exist in clusters with different
ages. However, the MS peak \textbf{barely appears} in young clusters
with age $\lesssim$ 1 Gyr; and it disperses gradually with
increasing age.
   \begin{figure}
     \includegraphics[angle=-90, width=7.5cm]{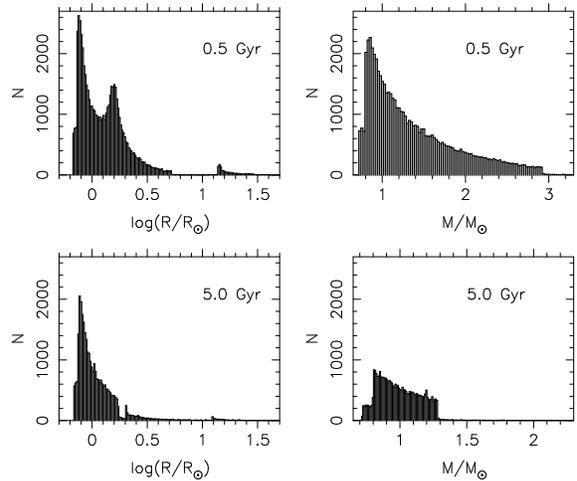}
     \centering
     \caption{The histograms of \textbf{the} radius and mass of the simulated
     clusters with Z = 0.02.}
     \label{figmr}
   \end{figure}

\textbf{A} RC peak had been observed by \emph{CoRoT} \citep{hekk09}.
When age $>$ 2.0 Gyr, the dominant RC peak is located
\textbf{approximately} between 2.5 and 4 $\mu$Hz, which is not
sensitive to age. However, when age $<$ 2 Gyr, the location of the
dominant RC peak is sensitive to age. For example, the dominant peak
is located at about 3.5-4.5 $\mu$Hz for the cluster with age = 0.5
Gyr, whereas it is located at about 6-7 $\mu$Hz for the cluster with
age = 1.0 Gyr. Moreover, our calculations show that the dominant RC
peak of \textbf{the} cluster with age $\simeq$ 1.2 Gyr is located at
\textbf{approximately} 7-8 \dhz{}, which is higher than the values
of other clusters. The RC peak is caused by \textbf{the fact} that
the radius of many RC stars near the ZAHB concentrates in a certain
range. For example, the mass of RC stars of the simulated cluster
with Z =0.02 and age = 1.0 Gyr is almost uniformly distributed in
the range 2.13-2.28 \dsunm{}. However, most \textbf{of} these stars
have a radius in the range $\sim$ 9.3-10.5 $R_{\odot}$, which
results in a concentration of \ddel{} in the range \textbf{of} about
6-7 \dhz{}.

In stars more massive than about 2.0 \dsunm{} (this value can be
affected by the convective overshooting parameter $\delta_{ov}$),
helium-burning temperatures are reached at the center before
electrons become degenerate there \citep{hurl00}. For these stars,
when they reach the `ZAHB', their mean densities decrease with
increasing mass, just \textbf{like} ZAMS. However, for stars with $M
<$ 2.0 \dsunm{}, electrons in the hydrogen-exhausted core are highly
degenerate before helium ignition occurs. The radius of these stars
on the ZAHB depends mainly on the mass of the hydrogen-exhausted
core at the time of helium ignition and on the mass \textbf{of} the
overlying envelope. The mean density of these stars on the ZAHB does
not \textbf{depend simply} on the total mass, but it is usually less
than \textbf{the mean density} of stars with $M \simeq$ 2.0
\dsunm{}. Thus \textbf{a} ZAHB star with $M \simeq$ 2.0 \dsunm{} has
the highest mean density. Furthermore, for stars with $M <$ 2.0
\dsunm{} the temperature and density distributions in the core are
\textbf{almost} unique functions of the core mass \citep{iben74}.
For stars with $M < $ 1.6 \dsunm{}, when they reach the ZAHB,
\textbf{their radius and the mass of their envelopes} decrease with
the decrease in the total mass, but the core mass is approximately
equal. Their mean densities are approximately equal too, which leads
to a similar value of \ddel{} for these stars on the ZAHB. The mass
of many ZAHB stars in \textbf{the} cluster with Z = 0.02 and age
$\simeq$ 1.2 Gyr is about 2.0 \dsunm{}, hence the frequency of the
dominant RC peak of this cluster is larger than that of other
clusters. The mass of RC stars \textbf{in} clusters with Z = 0.02
and age $>$ 2.4 Gyr is less than 1.6 \dsunm{}, so that the locations
of the dominant RC \textbf{peaks} of these clusters are almost same.
The asteroseismical observation for red giant stars of clusters with
age $\lesssim$ 2.4 Gyr may provide a help to test the theory of the
degeneracy of hydrogen-exhausted core. Differing from the MS peak
that \textbf{exists only} in young clusters, the RC peak
\textbf{exists in almost} all clusters. This is because the RC stars
near the ZAHB exist in clusters with different ages in our
simulations.

The histograms of \textbf{the} radius and mass of clusters with age
= 0.5 and 5.0 Gyr are shown in Fig. \ref{figmr}. The
\textbf{distributions of the} mass \textbf{have} only one peak, at
about 0.8 \dsunm{}, which is caused by IMF. However, the
distributions of the radius are similar to those of \dnu{} and
\ddel{}. There are three peaks in \textbf{radius distribution} of
\textbf{the} cluster with age = 0.5 Gyr. The `low-radius' peak
\textbf{corresponds} to the HF peak of \ddel{}, the `middle-radius'
peak \textbf{to} the MS peak of \ddel{}, and the `high-radius' peak
\textbf{to} the RC peak of \ddel{}. However, when age = 5 Gyr, there
is no `middle-radius' peak in the distribution of \textbf{the}
radius and also no MS peak in the distribution of \ddel{}. This
clearly shows that the MS peak is caused by \textbf{the fact that}
many young MS stars \textbf{have approximately the same radii}, and
the distribution of \ddel{} is more sensitive to the distribution of
\textbf{the} radius than to that of \textbf{the} mass.

Moreover, Fig. \ref{figmr} shows that both the `low-mass' peak and
`low-radius' peak are significant for the cluster with Z = 0.02 and
age = 0.5 Gyr. They \textbf{correspond} to the HF peaks. However,
Fig. \ref{fig1} shows that the MS peaks of \dnu{} and \ddel{}
\textbf{are} more significant than the HF peaks. This is because the
\ddel{} of `low-mass' stars decreases more quickly with increasing
mass than that of `middle-mass' stars in the cluster. For example,
when the mass of a star with Z = 0.02 increases from 0.8 to 0.9
\dsunm{}, its radius increases from \textbf{approximately} 0.727 to
0.806 $R_{\odot}$ at 0.5 Gyr, which leads to a decrease of about
17.5 \dhz{} in \textbf{the} \ddel{} of \textbf{that} star. However,
when the mass of a star increases from 1.55 to 1.65 \dsunm{}, its
radius increases from 1.566 to 1.644 $R_{\odot}$, which results in a
decrease of about \textbf{only} 3.5 \dhz{} in \textbf{the} \ddel{}
of \textbf{that} star. Thus, although the number of stars with mass
between 0.8 and 0.9 \dsunm{} is about twice as much as that of stars
with mass between 1.55 and 1.65 \dsunm{}, the interval of \ddel{} of
the former is 5 times as wide as that of the latter. Therefore the
MS peaks are more significant than the HF peaks.

Using Eggleton's stellar evolution code, we calculated the
evolutions of stars with Z = 0.02 and mass between 1.55 and 1.65
\dsunm{}. Our calculations show that these stars have an approximate
radius between 1.465 and 1.492 $\mathrm{R}_{\odot}$ at the age of
about 40 Myr, which leads to almost \textbf{the} same mean density,
i.e. almost \textbf{the} same \ddel{}. However, the mean density of
a MS star decreases with age. Since a high mass star evolves faster
than a lower mass star, the mean density of \textbf{a} star with $M$
= 1.65 \dsunm{} decreases faster than that of \textbf{a} star with
$M$ = 1.55 \dsunm{}. Thus the \ddel{} of stars with mass between
1.55 and 1.65 \dsunm{} disperses gradually with increasing age. For
example, the difference \textbf{in} \ddel{} between these two models
is about 3.5 \dhz{} at the age of 0.5 Gyr, but it is about 8 \dhz{}
at the age of 1.0 Gyr. Thus \textbf{a} MS peak exists only in young
clusters. It moves toward a low frequency and disperses gradually
with increasing age.

Figs. \ref{fig1} and \ref{fig3} show that there is another peak on
the left of the MS gap for \textbf{the} cluster with age = 5.0 Gyr.
For stars with $M <$ 1.3 \dsunm{}, when their central convection
ceases, the central hydrogen is not exhausted. The lower the mass,
the more the \textbf{remaining} hydrogen. Thus these stars can stay
between \textbf{the} MS hook and Hertzsprung gap for a long time,
which leads to the presence of a peak or bump between the RC-MS gap
and the MS gap. For the cluster with Z = 0.02 and age = 7.0 Gy, the
peak can become a bump. For stars with $M >$ 1.3 \dsunm{}, when the
central convection ceases, the central hydrogen is almost exhausted.
After the MS hook, they evolve rapidly into the Hertzsprung gap. In
addition, the higher the mass, the shorter the thermal time-scale.
Therefore the number of stars in the MS gap of a `young' cluster is
less than that in the MS gap of an `old' cluster, and there is no
the peak between the MS gap and RC-MS gap in `young' clusters. The
MS hook takes place after the central hydrogen abundance of stars
decreases to about 0.05. The central hydrogen abundance of stars in
the MS gap is about 0.05-0.001, but that of stars in the peak on the
left of the MS gap is less than 0.001. For clusters with $M_{hook}
>$ 1.3 \dsunm{}, the central hydrogen abundance of most stars in the
lower boundary of \textbf{the} histogram of \textbf{the} \ddel{} of
MS stars should be about 0.05. Thus the asteroseismical observation
for turn-off stars of `middle-age' clusters is very important for
testing theories of stellar evolution and stellar population
synthesis.

\begin{figure}
     \includegraphics[angle=-90, width=7.5cm]{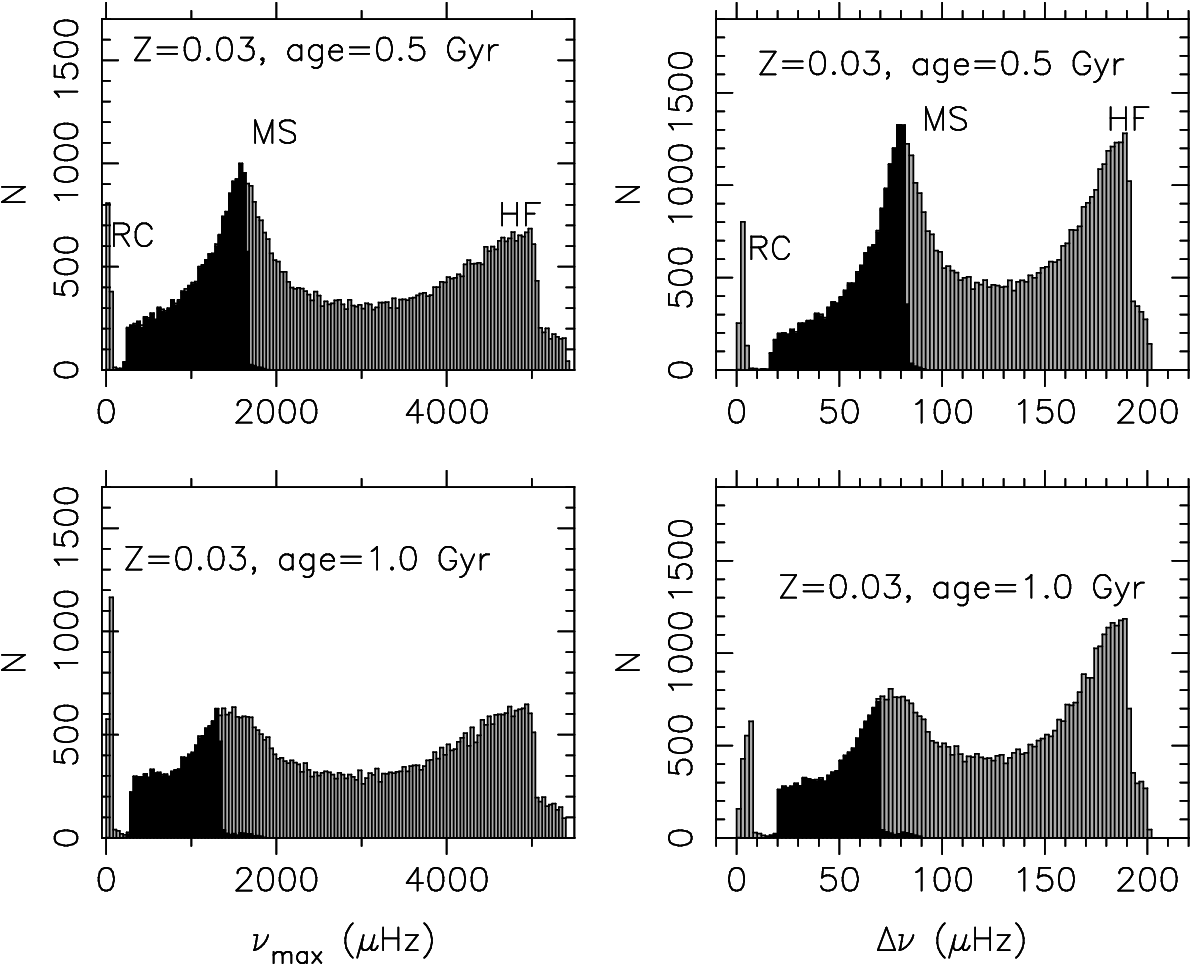}
     \includegraphics[angle=-90, width=7.5cm]{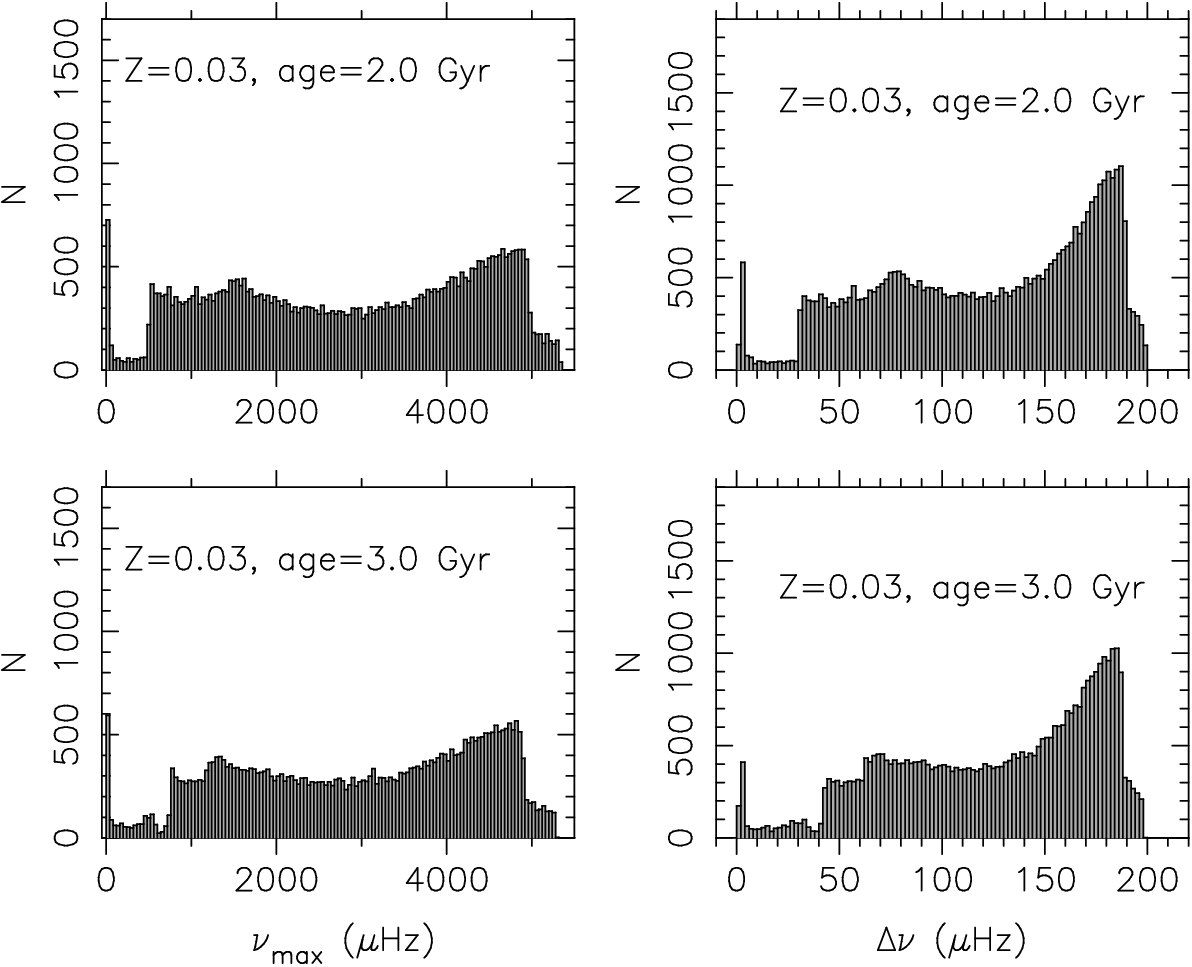}
     \includegraphics[angle=-90, width=7.5cm]{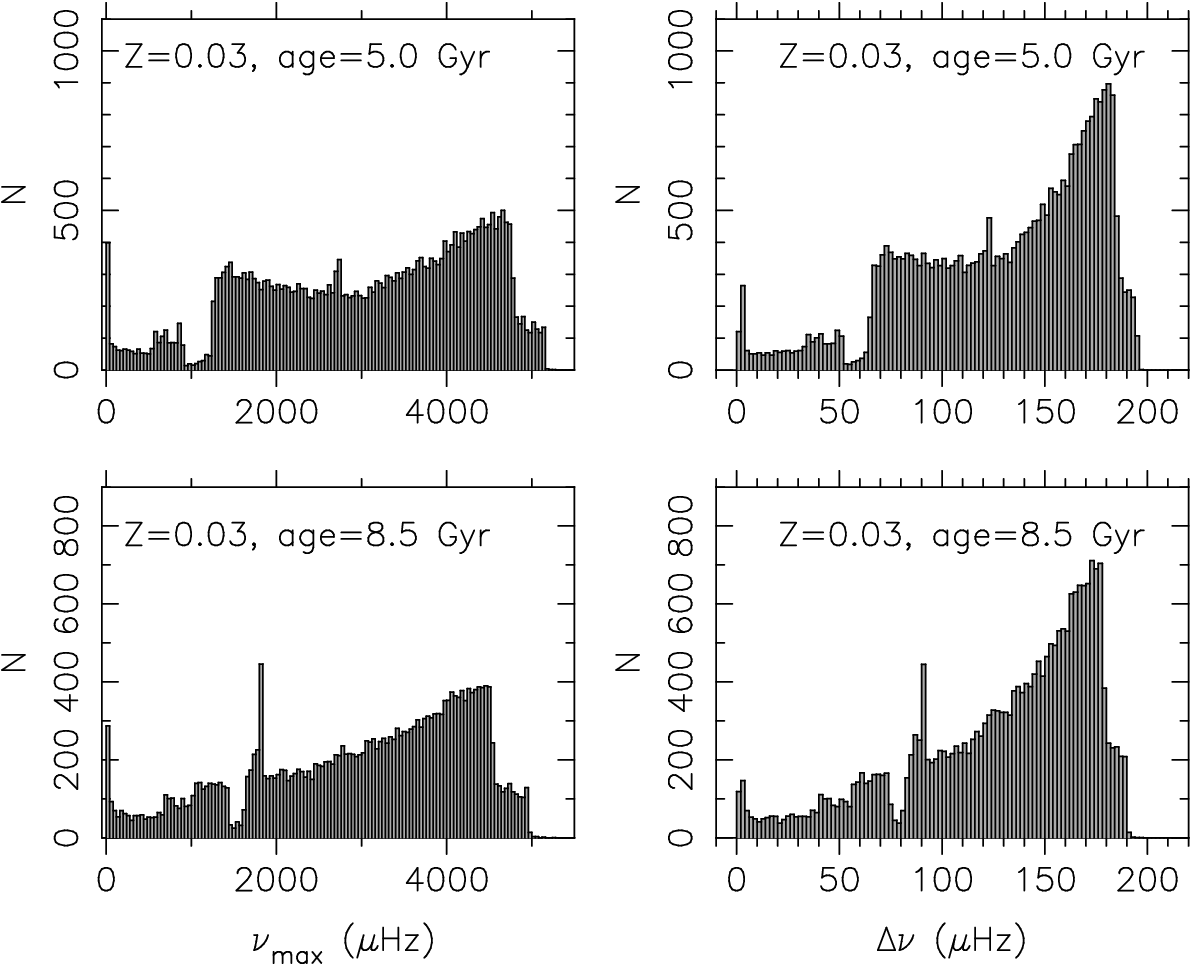}
     \centering
     \caption{Same as Fig.\ref{fig1} but for Z = 0.03.}
       \label{pz3}
   \end{figure}

Moreover, Figs. \ref{fig1} and \ref{pz3} show that there is a sharp
peak in the distributions of \textbf{the} \dnu{} and \ddel{} of MS
stars of clusters with age = 5.0 and 8.5 Gyr. When age increases
from 5.0 to 8.5 Gyr for clusters with Z = 0.02, the location of the
sharp peak moves from about 2860 to around 1920 \dhz{} for \dnu{}
and shifts from about 130 to around 95 \dhz{} for \ddel{}. These
peaks also exist in the distributions of \dnu{} and \ddel{} of the
cluster with Z = 0.008 and age = 7.0 Gyr. \textbf{These sharp peaks}
are \textbf{derived} from the models with mass between
\textbf{approximately} 1.02 and 1.04 \dsunm{} for all clusters. The
mean density of the models decreases with age. Thus, the locations
of the sharp peaks move from a high frequency to a lower one. Using
Eggleton's stellar evolution code, \textbf{we performed} detailed
model calculations. \textbf{We did not} find any particularity in
the evolutions of the models \textbf{with mass between 1.02 and 1.04
\dsunm{}}. Comparing evolutionary tracks computed using Eggleton's
code \textbf{with those computed using the Hurley code}, we find
that the Hurley code cannot reproduce exactly the evolutionary
tracks of late phases of \textbf{the} MS of the models with mass
between 1.02 and 1.04 \dsunm{}. \textbf{Nor can it reproduce well
the} mean densities of the models. \textbf{According to the Hurley
code, the mean densities} are almost equal, which leads to the
appearance of the sharp peaks. In fact, the mean densities of the
models \textbf{as} given by Eggleton's code are not equal.
\textbf{In any case}, the sharp peaks \textbf{that appear in the
application of the Hurley code} are artefacts. \textbf{Because} the
Hurley code \textbf{fails to reproduce well the mean densities of
these models, it} in effect \textbf{places the} models \textbf{one}
on top of another in the \dnu{} and \ddel{} plots.
\textbf{Nonetheless, the Hurley code can reproduce well the} mean
density as well as the evolutionary tracks of early phases of
\textbf{the} MS of the models with mass between about 1.02 and 1.04
\dsunm{}. Thus the sharp peaks do not appear in the distributions of
\dnu{} and \ddel{} of clusters with age $\leq$ 3.0 Gyr.

   \begin{figure}
     \includegraphics[angle=-90, width=7.5cm]{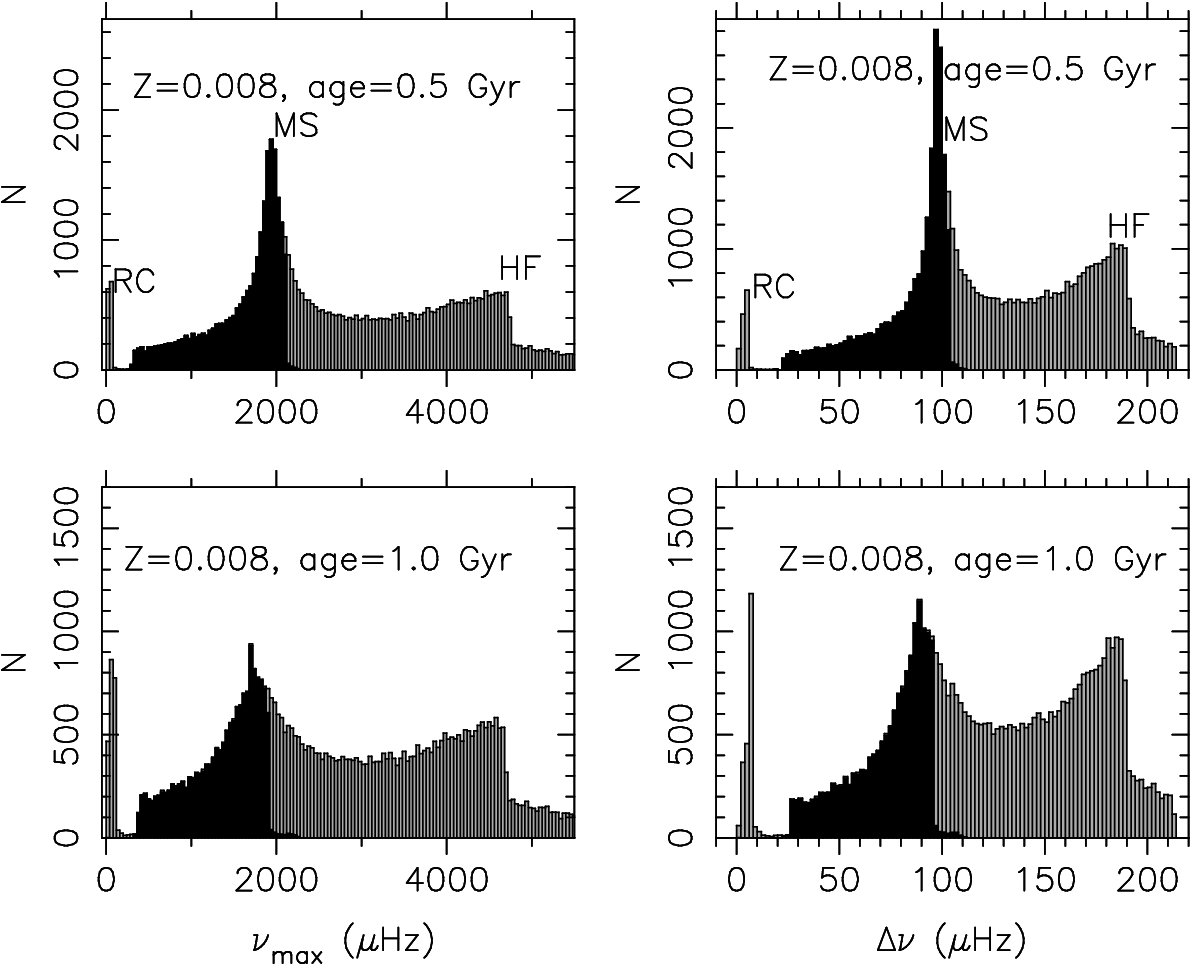}
     \includegraphics[angle=-90, width=7.5cm]{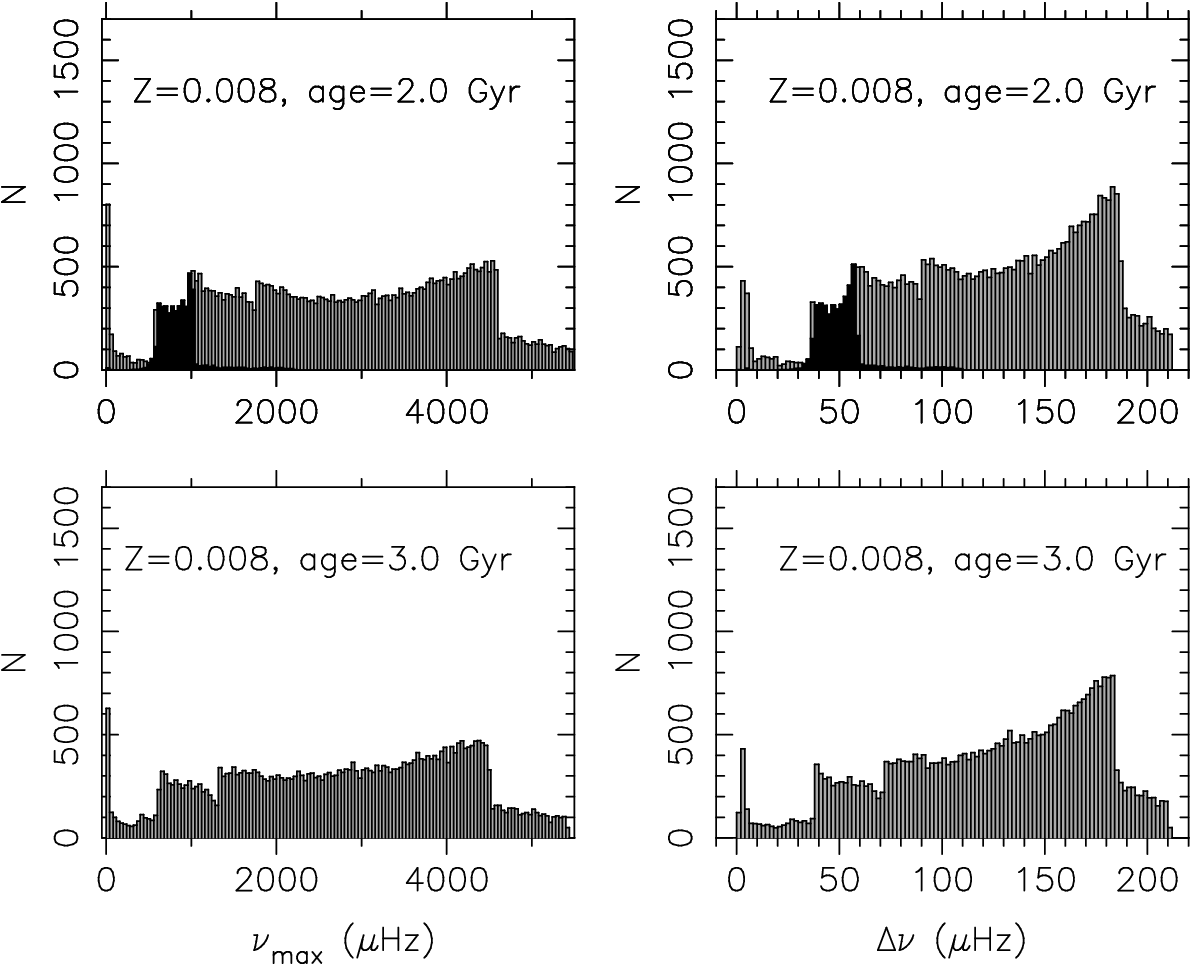}
     \includegraphics[angle=-90, width=7.5cm]{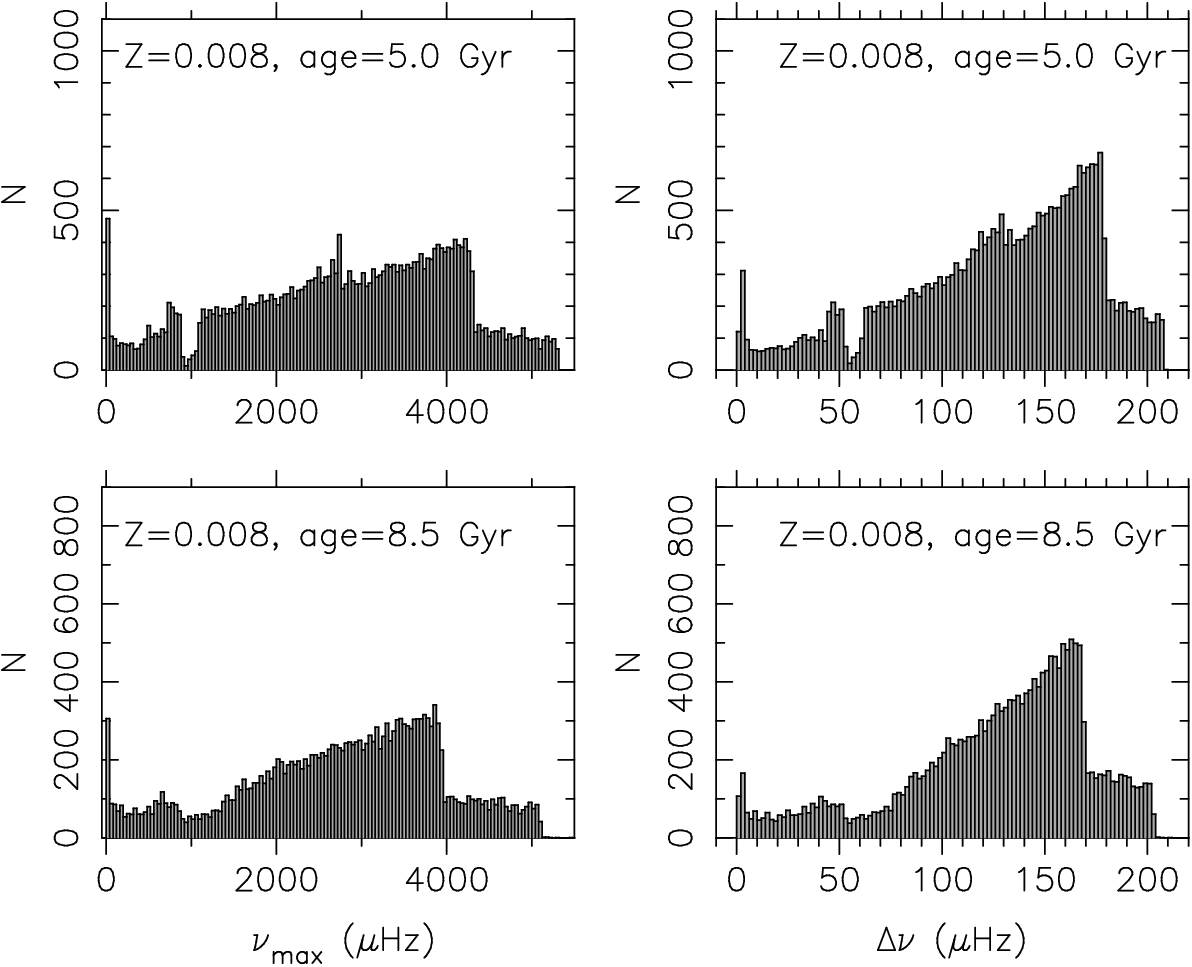}
     \centering
     \caption{Same as Fig.\ref{fig1} but for Z = 0.008.}
     \label{pz08}
   \end{figure}
   \begin{figure}
     \includegraphics[angle=-90, width=7.5cm]{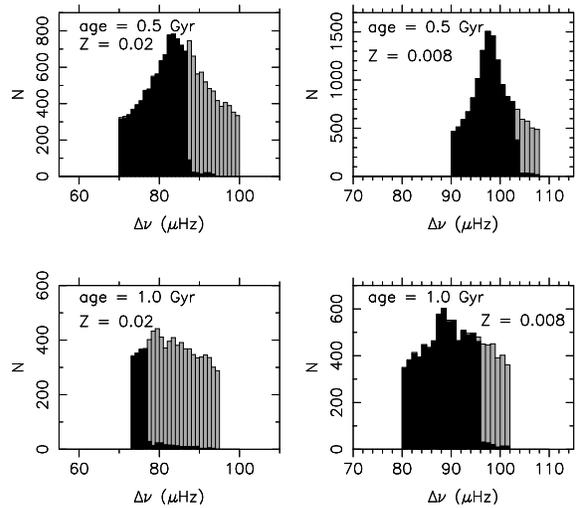}
     \centering
     \caption{The MS peaks of clusters with different
     ages and metallicities.}
     \label{fmsp}
   \end{figure}

\subsection{Effect of metallicity on peak locations}

The distributions of \dnu{} and \ddel{} of clusters with Z = 0.03
and 0.008 are shown in Figs. \ref{pz3} and \ref{pz08}, respectively.
For young clusters, \textbf{these distributions} have similar
characteristics: a gap and three peaks. Figs. \ref{fig1}, \ref{pz3}
and \ref{pz08} show that the distributions of \dnu{} and \ddel{} are
\textbf{obviously affected} by metallicity. In spite of the fact
that there are many stars hotter than the red edge of the
instability strip, Fig. \ref{fmsp} shows that the location of
\textbf{the} dominant MS peak is \textbf{sensitive not only} to age,
but also to metallicity. When age increases from 0.5 Gyr to 1.0 Gyr,
the frequency of the peak location decreases from 83 $\mu$Hz to 79
$\mu$Hz for clusters with Z = 0.02, but decreases from 98 $\mu$Hz to
88 $\mu$Hz for clusters with Z = 0.008. However, the location of
dominant RC peak is not sensitive to metallicity. For example, for
clusters with age = 1.0 Gyr, the peak location shifts only $\sim$ 1
\dhz{} when the metallicity decreases from 0.03 to 0.008. In young
clusters with age $\lesssim$ 2.0 Gyr, there are many stars hotter
than the red edge of the instability strip. \textbf{In a} survey
\textbf{of} solar-like oscillations, \textbf{these stars} may be
discarded. \textbf{The location of the} MS peak \textbf{would then}
be affected. For the cluster with Z = 0.008 and age = 0.5 Gyr, the
MS peak is mainly composed of \textbf{such} stars; \textbf{so there
should exist no} MS peak in the distribution of solar-like
oscillation frequencies. For clusters with Z $\gtrsim$ 0.02,
\textbf{however,} the stars hotter than the red edge of the
instability strip are \textbf{located mainly} on the left of the
dominant MS peaks of \dnu{} and \ddel{}. Hence \textbf{a} MS peak
may exist in these clusters. The asteroseismical observation of
young clusters with a high metallicity, especially for MS stars, may
aid in determining the age and metallicity of \textbf{these}
clusters.

\section{Discussions and Conclusions}

In our simulations, we calculated the evolutions of 5 $\times
10^{4}$ binaries for a cluster. However, even \textbf{though} the
number of initial models decreases to 2000 binaries, our results
\textbf{are not} affected. \textbf{Stars} hotter than the red edge
of the instability strip are not discarded in our models\textbf{,
but} they have been shown in black. For \textbf{all} clusters with
age $\gtrsim$ 2.0 Gyr except the one with Z = 0.008, the
distributions of \dnu{} and \ddel{} are not affected by these stars.
Although these stars are not expected to exhibit solar-like
oscillations, they may exhibit other \textbf{kinds of oscillation}.

The evolutions of wide binary stars are similar to those of single
stars. Strong binary interactions, such as mass transfer and mass
accretion, can lead to \textbf{either} a decrease or increase in the
mass of \textbf{such binary} stars. The \dnu{} and \ddel{} values of
these stars can \textbf{obviously be changed}. Thus for a specific
binary system, the effect of binary interactions on oscillations is
significant. \textbf{On} the one hand, \textbf{however,} the
fraction of these interactive binary stars in our samples is very
small. On the other hand, the effect of mass loss of some stars on
the distributions of \dnu{} and \ddel{} can be partly counteracted
by that of mass accretion of other stars. Therefore, the effect of
binary interactions on the distributions of \dnu{} and \ddel{} is
not significant. We also calculated the BSP using \cite{salp55} IMF
and other assumptions for $q$, $e$ and $a$, such as $n(q) = 2q$,
$n(e) = 2e$ and $n(\mathrm{log}(a))$ = constant \citep{hurl02}, and
single-star stellar population. Our calculations show that results
are similar. Thus the assumptions concerning the BSP can not
significantly affect the distributions of \dnu{} and \ddel{}
\textbf{for} the fraction of interactive binary stars is not
\textbf{significantly enhanced}.

The OPAL \citep{roge92} used in the Eggleton code is out of date.
However, \cite{chen07} show that the effect of the differences
between OPAL \citep{roge92} and OPAL \citep{igle96} on the
evolutions of low-mass stars is negligible. Element diffusion for
helium and metals is very important in the evolution of solar-type
stars, especially for asteroseismology. However, this effect cannot
be considered in the Hurley code. If this effect is considered, our
results may be affected. The distributions of \dnu{} and \ddel{} of
RC stars can be affected by the mixing-length parameter $\alpha$
\citep{yang10b}. Moreover, the distributions can be affected by the
overshooting parameter $\delta_{ov}$. The RC peak moves to a lower
frequency when the $\delta_{ov}$ increases.

For the simulated clusters with a given age, the lower the
metallicity, the lower the mass of stars in the same evolutionary
phase. Thus the locations of the RC and MS peaks move to a higher
frequency when metallicity decreases. Furthermore, the radius of MS
stars is more sensitive to the initial metallicity than \textbf{the
radius} of RC stats; \textbf{so the} MS peak is more sensitive
\textbf{than RC peak to} the metallicity.

The age of the open clusters NGC 6866, NGC 6811, NGC 6819 and NGC
6791 is \textbf{approximately} 0.4, 1.0, 2.5 and 8.5 Gyr
\citep{stel10b}, respectively. The fraction of FGB and RC stars in a
cluster is scarcely affected by metallicity. \textbf{According to
our calculations,} about 90 per cent of red giants of NGC 6866 and
NGC 6811 could be RC stars, \textbf{but} about 80 per cent of red
giants of NGC 6791 could be FGB stars. \textbf{In} cluster NGC 6819,
the fraction of RC stars could be approximately equal to that of FGB
stars. For the cluster NGC 6791, because the values of \textbf{the}
\ddel{} of most FGB stars \textbf{are} larger than those of RC
stars, most FGB stars could be distinguished by asteroseismical
observation. The MS peak of \ddel{} might be observed at about 85-95
\dhz{} in NGC 6866, and the dominant peak of RC stars could be
located between 6 and 8 \dhz{} for NGC 6811.

For young clusters, there are three peaks and a gap in the
distributions of \dnu{} and \ddel{}. The gap \textbf{corresponds} to
the Hertzsprung gap in the Hertzsprung-Russell diagram. The MS peak
exists only in `young-age' clusters. The dominant MS peak moves to a
lower frequency with increasing age or metallicity. \textbf{Stars}
hotter than the red edge of the instability strip are almost
\textbf{all} located on the left of the dominant MS peak for
clusters with Z $\gtrsim$ 0.02. In addition, for clusters with
\textbf{1.1 \dsunm{} $<M_{hook}<$ 1.3 \dsunm{}}, there are a MS gap
and a peak on the left of the MS gap in the distributions \textbf{of
\dnu{} and \ddel{}}. The RC peak is caused by \textbf{the fact} that
the radius of many \textbf{RC stars near the} ZAHB
\textbf{concentrates} in a certain range; and the MS peak is caused
by \textbf{the fact} that the radius of many young MS stars with
mass in a certain range concentrates in a certain range. In other
words, \textbf{in a cluster, the mean density of many RC stars is
approximately the same, which leads to the RC peak; the mean density
of many young MS stars with mass in a certain range is approximately
the same, which leads to the MS peak.} For clusters with Z = 0.02,
the frequency of location of \textbf{the} dominant RC peak increases
with age when age $\lesssim$ 1.2 Gyr. \textbf{It} then decreases
with age when age $>$ 1.2 Gyr, but \textbf{it} scarcely varies when
age $>$ 2.4 Gyr. The helium core of stars with $M \gtrsim$ 2.0
\dsunm{} is non-degenerate, but that of stars with $M <$ 2.0
\dsunm{} is degenerate at the time of helium ignition, which results
in \textbf{the fact} that the change of \textbf{the} RC peak with
age is different. The dominant peak of RC stars of the cluster with
$M_{ZAHB} \simeq$ 2.0 \dsunm{} is located at the highest frequency.
The asteroseismical observation for clusters with age $\lesssim$ 2.4
Gyr may aid in testing the theory of the degeneracy of
hydrogen-exhausted core.

\section*{Acknowledgments}
We thank the anonymous referee for his/her helpful comments
\textbf{and Daniel Kister for his help}. This work was supported by
China Postdoctoral Science Foundation funded project 20100480222,
the Ministry of Science and Technology of the People's republic of
China through grant 2007CB815406, the NSFC though grants 10773003,
10933002, 10963001, 11003003, and project of the fundamental and
frontier research of Henan province under grant no. 102300410223.

\end{document}